# A delayed-choice thought-experiment with later-time entanglement

C. F. Boyle[1]
University of Central England, Birmingham B42 2SU, U.K.
and
R. L. Schafir[2]
CISM, London Metropolitan University, London EC3N 1JY, U.K.

## Abstract

It is possible to find the nonlocality type of correlations between particle pairs retrospectively, matched with the outcomes of a future entangling measurement. But this does not imply nonlocality in subensembles of product pairs, nor does it imply an influence propagating backwards in time.

Most delayed-choice thought-experiments have been obtained purely from the superposition property of quantum mechanics, though entanglement appeared in the last of Wheeler's seven examples of delayed-choice [1], and more recently Englert, Scully and Walther [2] incorporated entanglement into their version of the two-slit experiment as a method of obtaining measurement information. Of course entanglement is itself superposition applied to product states, but the mysteries which result from it, in particular nonlocality, seem to go beyond the purely superposition mysteries.

However it is possible to construct a delayed-choice situation in which the later choice is a measurement of entangled states, though as we will see, it brings the further mystery that it apparently creates nonlocality "in advance", from states which were simply product states at the time. But we will see how to resolve this problem.

Let us first look at a normal entanglement situation. Consider an ensemble of entangled particle pairs, which are disentangled by two spacelike-separated observers who find that their single-particle measurements have correlations from which we can deduce the existence of nonlocality [3]. For example, suppose we have polarized photons with states of vertical or horizontal polarization ($|V\rangle$ or $|H\rangle$), and each entangled pair, with an equal probability of ¼, is in one of the four Bell basis [4] states:

$$B_1, B_2 = \sqrt{½}[|V_1\rangle|V_2\rangle \pm |H_1\rangle|H_2\rangle]$$
$$B_3, B_4 = \sqrt{½}[|V_1\rangle|H_2\rangle \pm |H_1\rangle|V_2\rangle] \quad (1)$$

Then if the two spacelike-separated observers measure the particles at angles of $\theta$ and $\phi$, i.e. they measure in the product basis of $\theta$ and $\phi$ eigenstates:

[1] e-mail: conall.boyle@uce.ac.uk
[2] e-mail: roger.schafir @londonmet.ac.uk; shfeer@hotmail.com

$$(\cos\theta|V_1\rangle + \sin\theta|H_1\rangle)(\cos\phi|V_2\rangle + \sin\phi|H_2\rangle)$$
$$(\cos\theta|V_1\rangle + \sin\theta|H_1\rangle)(-\sin\phi|V_2\rangle + \cos\phi|H_2\rangle)$$
$$(-\sin\theta|V_1\rangle + \cos\theta|H_1\rangle)(\cos\phi|V_2\rangle + \sin\phi|H_2\rangle) \quad (2)$$
$$(-\sin\theta|V_1\rangle + \cos\theta|H_1\rangle)(-\sin\phi|V_2\rangle + \cos\phi|H_2\rangle)$$

we can obtain the usual result, by the usual calculations, that the particle pairs matched with each Bell basis outcome are correlated. For instance those pairs which had previously been in the Bell state $B_1$ have probability $\cos^2(\theta - \phi)$ of giving the same result for each observer (i.e. of both polarizations being along each observer's direction of measurement or both perpendicular to it) and probability $\sin^2(\theta - \phi)$ of being different, and therefore a correlation (the difference of these two probabilities) of $\cos 2(\theta - \phi)$.

However, if we calculate the probabilities and the correlation for the whole ensemble, irrespective of which Bell basis a pair had previously been in, we find probabilities of ½ of each of being the same and ½ of each being different, and so a correlation of zero. So although there is no objection in fundamental principle to the two observers (when they are able to come together afterwards and compare records) being able to deduce information from the correlations about which entangled state the pairs had been in, because this was in the timelike past of both of them at the time they made the measurements, nevertheless it is not the case that they can do so. Provided the previous entangled states are not known to them, their records only show uncorrelated polarizations, with equal probability of being along or perpendicular to their chosen directions of measurement.

This suggests that we would get the same results if the particles are *first* measured in a product basis, *then* move together and are entangled by being measured in an entangled basis.

This is indeed the case – indeed the calculations are the same, so let us carry them out for a more general situation. Suppose we have a non-entangled ensemble of particle pairs, with relative frequencies $\alpha_{ij}$ of each possible outcome when two spacelike-separated observers measure them in a product basis:

$$|e_i\rangle_1 |e_j\rangle_2 \quad \text{with } i, j, = 1, 2 \quad (3)$$

and which come together afterwards and are then entangled by being measured in an entangled basis

$$\sum_{ij} a_{Aij} |e_i\rangle_1 |e_j\rangle_2 \quad (A = 1,\ldots,4,\ i,j = 1,2) \quad (4)$$

with:

$$\sum_{ij} \overline{a}_{Aij} a_{Bij} = \delta_{AB}, \quad \sum_A \overline{a}_{Aij} a_{Ars} = \delta_{ir}\delta_{js} \quad \text{(i.e. } a_{Aij} \text{ is unitary)} \quad (5)$$

Then:

$$\text{prob}\left\{|e_i\rangle_1 |e_j\rangle_2 \ \& \ \sum_{rs} a_{Ars} |e_r\rangle_1 |e_s\rangle_2 \right\} = \alpha_{ij} |a_{Aij}|^2 \quad (6)$$

Therefore:

$$\text{prob}\left\{\sum_{ij} a_{Aij} |e_i\rangle |e_j\rangle\right\} = \sum_{ij} \alpha_{ij} |a_{Aij}|^2 \tag{7}$$

and therefore:

$$\text{prob}\left\{|e_i\rangle |e_j\rangle \,\Big|\, \sum_{rs} a_{Ars} |e_r\rangle |e_s\rangle\right\} = \frac{\alpha_{ij} |a_{Aij}|^2}{\sum_{rs} \alpha_{rs} |a_{Ars}|^2} \tag{8}$$

So, although no pattern can be discerned until the later measurement is made, when it has been made it is found that in each earlier subensemble which was matched with each later outcome, there exist the correlations associated with entangled states.

Now suppose we allow for an alternative, incompatible, measurement at the later time in place of the entangled basis measurement in (4), and allow the observer who makes the later measurement to choose between them. For example the alternative choice might be a different entangled basis, or measurement in a product basis, or even no measurement at all. This gives us a "delayed choice" experiment with the same basic feature as discussed by Englert, Scully and Walther [2]. That is, it is possible to find one pattern (a pattern of correlations in this case) among the already existing results if one of the choices is made at the later measurement, or an alternative pattern (or no pattern) if a different choice is made, but at most one pattern can be found in any actual case, because of the incompatibility of the choices.

Let us look at some other features of this example. First of all, like the previous examples of delayed-choice, it is not a backwards-in-time effect – or at least, not provably so, since there are no grounds for asserting that any of the earlier outcomes would have been different if the later choice was different. It is simply a matter of which class each individual outcome belongs to in a classification of the results when the later choice has been made.

Yet each class in the retrospective classification possesses the correlations which are usually considered to imply the existence of nonlocality. Does this mean that nonlocality already exists among the particle pairs of a mixture of *product* states, though not revealed until the later measurement?[3]

The answer is no, and to see why let us review how nonlocality is proved (either by a Bell inequality argument [3] or a Hardy non-inequality argument [6]). We suppose there is *locality* for both observers, i.e. each observer's outcomes depend only on the same observer's choice of which observable to measure. Then it is possible to associate certain outcomes for each observer which are independent of the other observer's choice of what to measure (leading to four fixed sequences for the Bell (4-angle) case, or certain single outcomes for the Hardy case). But the result of this is to violate the necessary correlations, provided we assume that the contrafactual outcomes would also have obeyed the correlations, just as the actual outcomes do. (This is part of Stapp's concept of "contrafactual definiteness" [7], without which non-locality would only apply to hidden-variable theories.) Therefore each observer's outcomes cannot be independent of the far-away observer's choices, and therefore there cannot be locality for both observers.

In the present case, however, there can be a different reason why the outcomes may not be fixed. We are choosing subsequences matched against each of the possible outcomes of the later entangled basis measurement, and if one or other of the ear lier observers had made a different choice of what to measure then it might have changed the later entangled outcome. Therefore we might not be comparing each actual subsequence with the same subsequence in the contrafactual case (same in the sense of the same choice of sequence index, e.g. if the actual subsequence is the 1st, 2nd, 4th, 7th..., we imagine it compared with the 1st, 2nd, 4th, 7th... in the contrafactual case), and therefore we are not comparing like with like in comparing the actual and contrafactual subsequences.

---

[3] This situation is not to be confused with the "nonlocality without entanglement" of Bennett *et al* [5].

The physical influence by which the earlier observers' choices affect the outcomes of the later measurement might still be regarded as baffling, but at least the effect is in the timelike future of the earlier choices, so is not subject to the same objection in principle that a superluminal influence is.

But what if the later entangling measurement is not made? The correlations are still there, presumably, even if never discovered, so how do we avoid concluding that there exists a "hidden" nonlocality in this case?

To answer this, note another feature of the usual nonlocality argument. At a certain part of the argument we assume that an observer's outcome which would have occurred under a certain contrafactual measurement choice of the same observer would, by locality, still have occurred if the other observer had also made a contrafactual measurement choice. This "second round" of counterfactuals, in which contrafactual reasoning is applied to already-contrafactual results, appears to be an irremovable part of all proofs of nonlocality.

Supposing, then, that the later entangling measurement is never made, treat it as the alternative choice to the choice that *is* made. Then *if* it had been made, i.e. in the contrafactual case, the contrafactual outcomes can be dependent on which earlier choices were made by the spacelike-separated observers. This is exactly the same sort of "second round" contrafactual reasoning as already occurs in the standard proofs of nonlocality.

None of this invalidates, of course, previous nonlocality proofs which assume that each pair has been entangled in the past and that the disentangling measurements come later. Here there is no correspondingly "acceptable" method of avoiding nonlocality. If a different choice by the disentangling observers changed the entangled state, that would be an influence travelling into the timelike past which is no improvement on a superluminal influence. A backwards-in-time influence could also have been used to explain the apparent nonlocality for product states which appears in the above example, but we have seen that an assumption like that is unnecessary.

**Note added**  Since the first version of this article, related work by Peres [8] has been brought to our attention.

## References

[1]  J. A. Wheeler, "The Past and Delayed Choice Double Slit Experiment," in *Mathematical Foundations of Quantum Theory*, ed. Marlow A. R. (Academic Press, New York, 1978).

[2]  B–G Englert, M. O. Scully and H. Walther, *Nature*, **351**, 111 (1991); *Am. J. Phys.* **67** (4), 325 (1999).

[3]     A. Peres, *Quantum Theory: Concepts and Methods*, (Kluwer, Dortrecht etc, 1993), Ch. 6.

[4]     S. L. Braunstein, A. Mann and M. Revzen, *Phys. Rev. Lett.*, **68** (22), 3259 (1992).

[5]     C. H. Bennett, D. P. DiVicenzo, C. A. Fuchs, T. Mor, E. Rains, P. R. Shor, J. A. Smolin and W. K. Wootters, *Phys. Rev. A.*, **59**, 1070 (1999).

[6]     L. Hardy, *Phys. Rev. Lett.*, **71** (11), 1665 (1993).

[7]     H. P. Stapp, *Phys. Rev. D*, **3**, 1303 (1971); *Lectures on Bell's theorem*, Trieste (1975).

[8]     A. Peres, quant-ph/9904042.